\documentclass[aps,prc,twocolumn,superscriptaddress,showpacs,floatfix]{revtex4} 
\usepackage{epsfig} 
\usepackage{amsmath} 
\usepackage{bm} 
 
\begin{document} 
 
\title{Elliptic flow of thermal dileptons in relativistic nuclear collisions} 
\author{Rupa Chatterjee} 
\affiliation{Variable Energy Cyclotron  
Centre, 1/AF Bidhan Nagar, Kolkata 700 064, India}  
\author{Dinesh K.~Srivastava} 
\affiliation{Variable Energy Cyclotron  
Centre, 1/AF Bidhan Nagar, Kolkata 700 064, India}                         
\author{Ulrich Heinz} 
\affiliation{Physics Department, The Ohio State University,  
             Columbus, OH 43210, USA}  
\author{Charles Gale} 
\affiliation{Department of Physics, McGill University, 3600 University 
Street, Montr\'eal, H3A 2T8, Canada}             
 
\date{\today} 
 
\begin{abstract} 
We calculate the transverse momentum and invariant mass dependence  
of elliptic flow of thermal dileptons for Au+Au collisions at the  
Relativistic Heavy Ion Collider. The system is described using  
hydrodynamics, with the assumption of formation of a thermalized  
quark-gluon plasma at some early time, followed by cooling through  
expansion, hadronization and decoupling. Dileptons are emitted  
throughout the expansion history: by annihilation of quarks and  
anti-quarks in the early quark-gluon plasma stage and through a  
set of hadronic reactions during the late hadronic stage. 
The resulting differential elliptic flow exhibits a rich structure, 
with different dilepton mass windows providing access to different 
stages of the expansion history. Elliptic flow measurements  
for dileptons, combined with those of hadrons and direct photons, are  
a powerful tool for mapping the time-evolution of heavy-ion collisions.  
\end{abstract} 
 
\pacs{25.75.-q,12.38.Mh} 
\maketitle 
 
%%%%%%%%%%%%%%%%%%%%%%%%%%%%%%%%%%%%%%%%%%%%%%%%%%%%%%%%%%%%%%%%%%% 
\section{Introduction} 
\label{sec1}
%%%%%%%%%%%%%%%%%%%%%%%%%%%%%%%%%%%%%%%%%%%%%%%%%%%%%%%%%%%%%%%%%%% 
 
Evidence has been mounting for some time that collisions between heavy  
nuclei at ultra-relativistic energies lead to formation of quark-gluon  
plasma (QGP). Strong confirmations arise from recent observations at 
the Relativistic Heavy Ion Collider (RHIC) of large anisotropic flow  
of all hadronic species \cite{v2_exp,v2_part,v2_theo,Teaney} and of a  
suppression of high-$p_T$ hadrons due to parton energy loss in the dense  
medium \cite{jetq_theo,jetq_exp}. Signatures of direct photon emission  
\cite{simon,guy,bms_phot,fms_phot}, as well as preliminary results on 
(excess) dilepton production in such collisions, indicative of a hot 
early state, have also started emerging \cite{phenix1}. 
 
The radial and elliptic flow of hadrons observed in relativistic  
heavy-ion collisions at RHIC strongly suggest that  
the quark-gluon plasma (QGP) created in these collisions acts like a  
strongly coupled plasma with almost perfect liquid behaviour \cite{SQGP}.  
This conclusion is based on the successful prediction of the hadron  
momentum distributions, in particular of their anisotropies in  
non-central collisions, by dynamical calculations which treat the  
expanding QGP as an ideal fluid \cite{v2_theo,Teaney,QGP3,reviews,Hirano,% 
HHKLN}. While no other equally successful model exists, one has to remain  
conscious of the fact that the new ``perfect liquidity'' paradigm is  
based on a model back-extrapolation of the measured data to the early  
stages of the collision which are not directly accessible with hadronic  
observables. There are strong arguments that this back-extrapolation is  
fairly unique \cite{SQGP} and hence that the above-mentioned qualitative  
conclusion is robust. On a quantitative level, however, the extraction  
from experimental data of the (small) QGP viscosity is presently hampered  
not only by the unavailability of consistent hydrodynamic codes for  
{\em viscous} relativistic fluids, but even more by uncertainties about  
the hydrodynamic effects of changes in the equation of state of the QGP  
matter \cite{Pasi} and about details of the initial conditions at the  
beginning of the hydrodynamic expansion stage \cite{Hirano,Lappi}. 
 
The next generation of RHIC experiments will aim at obtaining more  
quantitative information about the properties of the QGP and the  
subsequent hot hadronic matter, as well as on the process of  
hadronization itself. While more precise hadronic flow measurements, 
in particular of elliptic flow in non-central collisions, will further 
constrain the hydrodynamic evolution models and the QGP equation of 
state, it will be invaluable to have additionally measurements on 
electromagnetic radiation, i.e. direct photons and dileptons which are 
penetrating probes \cite{KG}, and access more directly the very early 
expansion stages \cite{Ruuskanen}. Elliptic flow is generated very early, 
via  the transformation of the initial spatial eccentricity of the nuclear  
overlap region into momentum anisotropies through the action of  
azimuthally anisotropic pressure gradients. With the passage of time,  
the pressure gradients equalize, and the growth of elliptic flow shuts  
itself off \cite{sorge}. Photons and dileptons, which escape from the 
expanding fireball without re-interaction, will be able to probe  
specifically this early stage where the flow anisotropy first develops. 
Measurements of the elliptic flow of thermal photons and lepton pairs  
could thus provide particularly clean constraints on the QGP equation of 
state, with fewer ambiguities arising from theoretical uncertainties about  
the hadronization process and late-stage hadronic rescattering dynamics  
than in the case of hadronic flow measurements \cite{HHKLN}.  
 
The azimuthal asymmetry of photons from jet-plasma interactions has 
been examined recently \cite{TGF}, and so has the elliptic flow of 
thermal photons \cite{CFHS}, calculated with a hydrodynamic model for 
the expansion of the collision fireball. Contrary to the monotonic 
rise with $p_T$ of the hydrodynamically predicted differential elliptic 
flow $v_2(p_T)$ of hadrons, the thermal photon elliptic flow was shown 
to {\em peak} at transverse momenta of 1-2 GeV/$c$, decreasing at higher 
$p_T$ as a reflection of the decreasing hydrodynamic flow anisotropies 
as one goes backwards in the expansion history of the collision fireball. 
In addition, the photon elliptic flow was seen to exhibit an interesting 
peak-valley structure at low transverse momenta, $p_T\sim 0.4-0.5$\,GeV/$c$,  
reflecting the dominance of different types of hadronic photon production  
mechanisms below and above this $p_T$ value \cite{CFHS}. In the present 
paper we complement the work of \cite{CFHS} with a study of thermal 
dilepton (i.e. {\em virtual} photon) elliptic flow, where the invariant 
mass of the  
lepton pair (photon virtuality) provides an additional continuously  
tunable parameter. As we will show, the combined dependences of the  
elliptic flow of dileptons on their transverse momentum $p_T$ and  
invariant mass $M$ provide a very rich landscape of structures which  
can be used to set observational windows on specific stages of the  
fireball expansion. Future studies will further complement the analysis  
presented here by directly measuring the space-time structure of the  
early collision fireball with two-photon correlations. By combining the  
momentum structure of the (virtual) photon emitting source from flow  
measurements with its space-time structure from Hanbury Brown-Twiss  
correlations one should be able to provide complete experimental  
characterization of the phase-space structure of the QGP during the  
early stages of the fireball expansion. 
 
%%%%%%%%%%%%%%%%%%%%%%%%%%%%%%%%%%%%%%%%%%%%%%%%%%%%%%%%%%%%%%%%%%% 
\section{Dilepton emission rates and spectra} 
\label{sec2}
%%%%%%%%%%%%%%%%%%%%%%%%%%%%%%%%%%%%%%%%%%%%%%%%%%%%%%%%%%%%%%%%%%% 
 
Dileptons, like photons are emitted from every stage of a heavy-ion  
collision, from the pre-equilibrium stage~\cite{GK}, the quark-gluon 
fluid \cite{Feinberg76,Shuryak78,McLT85,HK,KLM,AR,Markus,SMM}, and  
the late hadronic matter \cite{KKMM,KKMR,vdm,dip,rapp,yul}. An interesting  
outcome of dilepton emission studies has been the confirmation of medium  
modifications of the spectral properties of $\rho$ mesons \cite{rapp,na60}.  
Correlated charm and bottom decay provide another important source of  
dileptons, which will help estimate energy loss and the elliptic flow  
of heavy quarks \cite{charm}. It may be possible in principle to isolate  
them by determination of the decay vertices.  
  
In the present paper we focus on thermal emission of dileptons from the  
QGP and hadronic phases in a collectively expanding fireball. The dilepton 
spectrum is obtained \cite{KKMR} by integrating the thermal emission rate  
over the space-time history of the system. As a result, high-$M$ dileptons  
arise mostly from the hot early stage where hydrodynamic flow is weak but  
the spatial eccentricity of the source is large, whereas dileptons of lower  
mass are emitted when the temperature is low, the flow is strong and  
anisotropic, but the spatial eccentricity of the fireball has mostly  
disappeared. These generic expectations are, however, modulated by the 
fact that dilepton emission from the late hadronic phase is characterized 
by strong vector meson resonance peaks in the dilepton mass spectrum which 
leads to strong variations of the relative weight of dilepton emission from 
the QGP and hadron gas phases on and off these resonances. This leads to 
very strong and interesting structures in the $p_T$-integrated elliptic  
flow $v_2(M)$ as a function of dilepton mass which can be exploited for 
differential fireball microscopy.  
 
The dilepton momentum spectrum can be written as  
\begin{eqnarray} 
\label{eq1} 
  E \frac{dN_{\ell \bar\ell}}{dM^2\,d^3p} &=&  
  \frac{dN_{\ell \bar\ell}}{dM^2\, p_T dp_T\,dy\, d\phi}  
  \\ 
  &=& \int\bigl[(...)\exp\bigl(-p{\cdot}u(x)/T(x)\bigr)\bigr]\,d^4x\,, 
  \nonumber 
\end{eqnarray} 
where the quantity inside the square brackets indicates the thermal  
emission rates from the QGP or hadronic matter, with the thermal Boltzmann 
factor extracted for emphasis. We parametrize the dilepton (virtual photon)  
4-momentum as  
\begin{equation} 
\label{eq2} 
p^\mu{\,=\,}(M_T \cosh Y,p_T\cos\phi,p_T\sin\phi,M_T \sinh Y) 
\end{equation} 
and the flow velocity of the fireball fluid $u^\mu$ as 
\begin{equation} 
\label{eq3}  
u^\mu = \gamma_T \Bigl(\cosh \eta,v_x(x,y,\tau),v_y(x,y,\tau),\sinh\eta\Bigr)  
\end{equation} 
with 
\begin{equation} 
\gamma_T = (1{-}v_T^2)^{-1/2},\qquad v_T^2 = v_x^2{+}v_y^2. 
\end{equation} 
(We assume boost-invariant longitudinal expansion \cite{bj}.)  
We use coordinates $\tau,x,y,\eta$ with volume element  
$d^4x = \tau\, d\tau \, dx \, dy\, d\eta$ where  
$\tau{\,=\,}(t^2{-}z^2)^{1/2}$ is the longitudinal proper time and 
$\eta{\,=\,}\tanh^{-1}(z/t)$ is the space-time rapidity. The dilepton  
momentum is parametrized by its rapidity $Y$, its transverse momentum  
$p_T{\,=\,}(p_x^2+p_y^2)^{1/2}$, and its azimuthal emission angle  
$\phi$.  
 
The dilepton energy in the local fluid rest frame, which  
enters the Boltzmann factor in the thermal emission rate in the  
combination $p{\cdot}u/T$, is thus given by 
\begin{equation} 
\label{eq4} 
 p \cdot u = \gamma_T  
 \bigl[M_T \cosh (Y{-}\eta)- p_T v_T\cos(\phi{-}\phi_v)\bigr]\,, 
\end{equation} 
where $M_T=(M^2+p_T^2)^{1/2}$ is the transverse mass and $M$ the invariant  
mass of the dilepton, and $\phi_v{\,=\,}\tan^{-1}(v_y/v_x)$ is the  
azimuthal angle of the transverse flow vector. This expression shows  
that the azimuthal anisotropy ($\phi$-dependence) of the dilepton  
spectrum, conventionally characterized by its Fourier coefficients  
$v_n$ (where only even $n$ contribute at $Y{\,=\,}0$), 
\begin{eqnarray} 
\label{eq5} 
\frac{dN(b)}{dM^2 d^2p_T \, dY}& = & \frac{dN(b)}{dM^2\, 2 \pi p_T dp_T \, dY} 
\\ 
&&\times\,\Bigl[1+2 v_2(M,p_T,b) \cos(2 \phi)+\dots\Bigr], 
\nonumber 
\end{eqnarray} 
is controlled by an interplay between the collective flow anisotropy 
and the geometric deformation of the temperature field $T(x,y,\tau)$ 
(see \cite{CFHS} for more details). It vanishes in the absence of radial  
flow, $v_T{\,=\,}0$.  
 
We use the boost invariant hydrodynamic code AZHYDRO \cite{AZHYDRO} 
which has been used extensively and successfully to describe midrapidity 
hadron production at RHIC and more recently to predict the elliptic flow of 
thermal photons \cite{CFHS}. We study Au+Au collisions at  
$\sqrt{s}{\,=\,}200\,A$\,GeV and use the same initial conditions as in 
\cite{CFHS}, namely an initial time $\tau_0{\,=\,}0.2$\,fm/$c$ for the 
beginning of the hydrodynamic stage, with an initial peak entropy 
density in central collisions of $s_0{\,=\,}351\,\mathrm{fm}^{-3}$, 
corresponding to a peak initial temperature of $T_0{\,=\,}520$\,MeV. 
Although the matter may not yet be completely thermalized at such an  
early time, we choose $\tau_0$ so small in order to also account at  
least partially for pre-equilibrium dilepton production at very early  
times \cite{GK,Seibert}. This contribution to the dilepton spectrum is  
important at large $M$ and and large $p_T$, and it will suppress  
$v_2^{\ell\bar\ell}$ there because very little transverse flow develops  
before about 0.5\,fm/$c$. The initial transverse entropy density profile  
is computed from a Glauber model assuming 75\% wounded nucleon and 25\%  
binary collision scaling of the initial entropy production \cite{QGP3}. 
This reproduces the collision centrality dependence of hadron production  
in Au+Au collisions at RHIC \cite{QGP3,KH05}. 
 
We assume that a thermally and chemically equilibrated plasma is formed  
at the initial time, and use the Born term for the production of dileptons  
from the QGP. It is known \cite{AR,Markus} that for thermal dileptons the  
corrections to the Born term are $\sim (2\pi\alpha_s/3) T^2/M^2$ and thus  
decrease rapidly with increasing dilepton mass. For the production of  
dileptons from the hadronic reactions, we consider the comprehensive  
set of hadronic reactions analyzed by Kvasnikova et al. \cite{yul},  
which are known to correctly reproduce the spectral density measured  
in $e^+e^-$ scattering. In \cite{yul} these rates are conveniently  
parametrized in the form $F_{\text{eff}}(M)\exp(-E/T)$. They have been 
shown to provide a good description of intermediate mass dilepton data  
from Pb+Pb collisions at the SPS \cite{yul}. However, since these rates 
are used here for lower invariant masses, some comments are in order. For 
instance, it is known that vector meson spectral densities are modified 
in the medium \cite{KG}, and such effects are not included in the rates
of Ref.~\cite{yul}. Still, as long as they do not produce severe 
distortions of the transverse momentum spectra, the dilepton elliptic 
flow signal evaluated here should be relatively robust.  
 
% 
%%%%%%%%%%%%%%%%%%%%%%%%%%%%%%%%%%%%% Fig. 1 %%%%%%%%%%%%%%%%%%%%%%%%%%%%%%% 
\begin{figure}[t] 
%\centerline{\epsfig{file=dndm2_0.ps,bb=51  208  554  575,width=7.9cm}} 
%\centerline{\epsfig{file=dndm2_dt_0new.ps,width=8.2cm}} 
\centerline{\epsfig{file=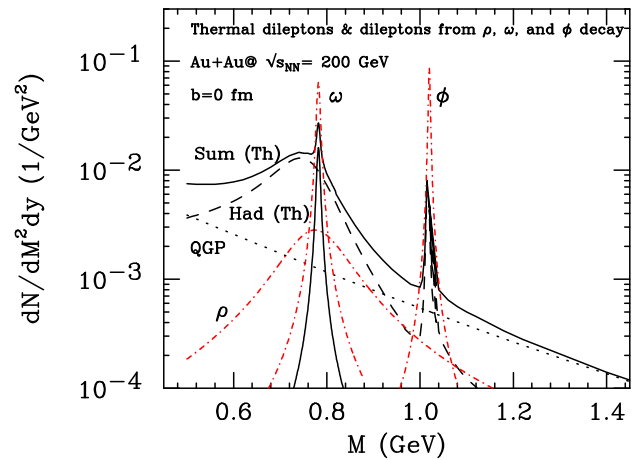,width=8.2cm}} 
\caption{(Color online)  
The mass spectrum of thermal dileptons from a hydrodynamical 
simulation of central 200\,$A$\,GeV Au+Au collisions ($b{\,=\,}0$). The  
quark and hadronic matter contributions are shown separately.  
See text for details. 
\label{fig1}} 
\end{figure} 
 
%%%%%%%%%%%%%%%%%%%%%%%%%%%%%%%%%%%%%%%%%%%%%%%%%%%%%%%%%%%%%%%%%%%%%%%%%%%% 
% 
 
For the equation of state we use EOS\,Q \cite{AZHYDRO} which matches a  
free quark-gluon gas (QGP) to a chemically equilibrated hadron resonance  
gas (HG) by a Maxwell construction at critical temperature  
$T_c{\,=\,}164$\,MeV, with energy densities  
$\epsilon_\mathrm{q}{\,=\,}1.6$\,GeV/fm$^3$ and  
$\epsilon_\mathrm{h}{\,=\,}0.45$\,GeV/fm$^3$ 
in the QGP and HG subphases at this temperature. Hadron freeze-out 
is assumed to happen at $\epsilon_\mathrm{f}{\,=\,}0.075$\,GeV/fm$^3$  
\cite{QGP3}. 
 
Figure~\ref{fig1} shows the production of thermal dileptons from 
the quark matter and the hadronic matter for central collisions.  
Similar figures emerge at other impact parameters and help us understand  
the relative importance of hadronic and quark-matter contributions 
at a given $M$. The QGP contribution (dotted line) is seen to completely  
dominate the mass spectrum of thermal dilepton radiation for dilepton  
masses above 1 GeV, except for the $\phi$ meson peak. Hadronic radiation 
(dashed line) dominates for $0.5\,\mathrm{GeV}<M<1\,\mathrm{GeV}$. 
%, but 
%for masses below 0.5\,GeV the QGP again outshines the hadron resonance  
%gas.  
The red dash-dotted lines show post-freeze-out decay contributions  
from $\rho$, $\omega$, and $\phi$ mesons; they are not included in the  
solid line for the total hydrodynamic spectrum. The sharply peaked solid  
black line under the $\omega$ peak (which {\em is} included in the total  
hydrodynamic spectrum) shows the contribution from thermally distributed  
$\omega$ mesons in the hadronic phase before freeze-out. Since the  
hadronic dilepton emission rates given in \cite{yul} do not include %any  
3-body scattering channels proceeding through the $\omega$ meson, we  
include the $\omega$ explicitly as a thermally distributed particle  
species, with its vacuum spectral function, integrating its density 
multiplied with the standard partial width for dilepton decay over the  
space-time volume of the hadronic phase contained inside the hadronic  
freeze-out surface. We see that post-freeze-out decays of $\omega$ 
and $\phi$ produce many more dileptons at the respective vector meson 
peaks than collisions and decays before hadronic freeze-out; this reflects  
the relatively long lifetimes of these two vector mesons (23.4 and  
44.5 fm/$c$, respectively). 
   
% 
%%%%%%%%%%%%%%%%%%%%%%%%%%%%%%%%%%%%% Fig. 2 %%%%%%%%%%%%%%%%%%%%%%%%%%%%%%% 
\begin{figure}[t] 
%\centerline{\epsfig{file=dn_rho.ps,bb=51  208  546  573,width=7.9cm}} 
%\centerline{\epsfig{file=v2_rho.ps,bb=54  208  546  568,width=7.9cm}} 
\centerline{\epsfig{file=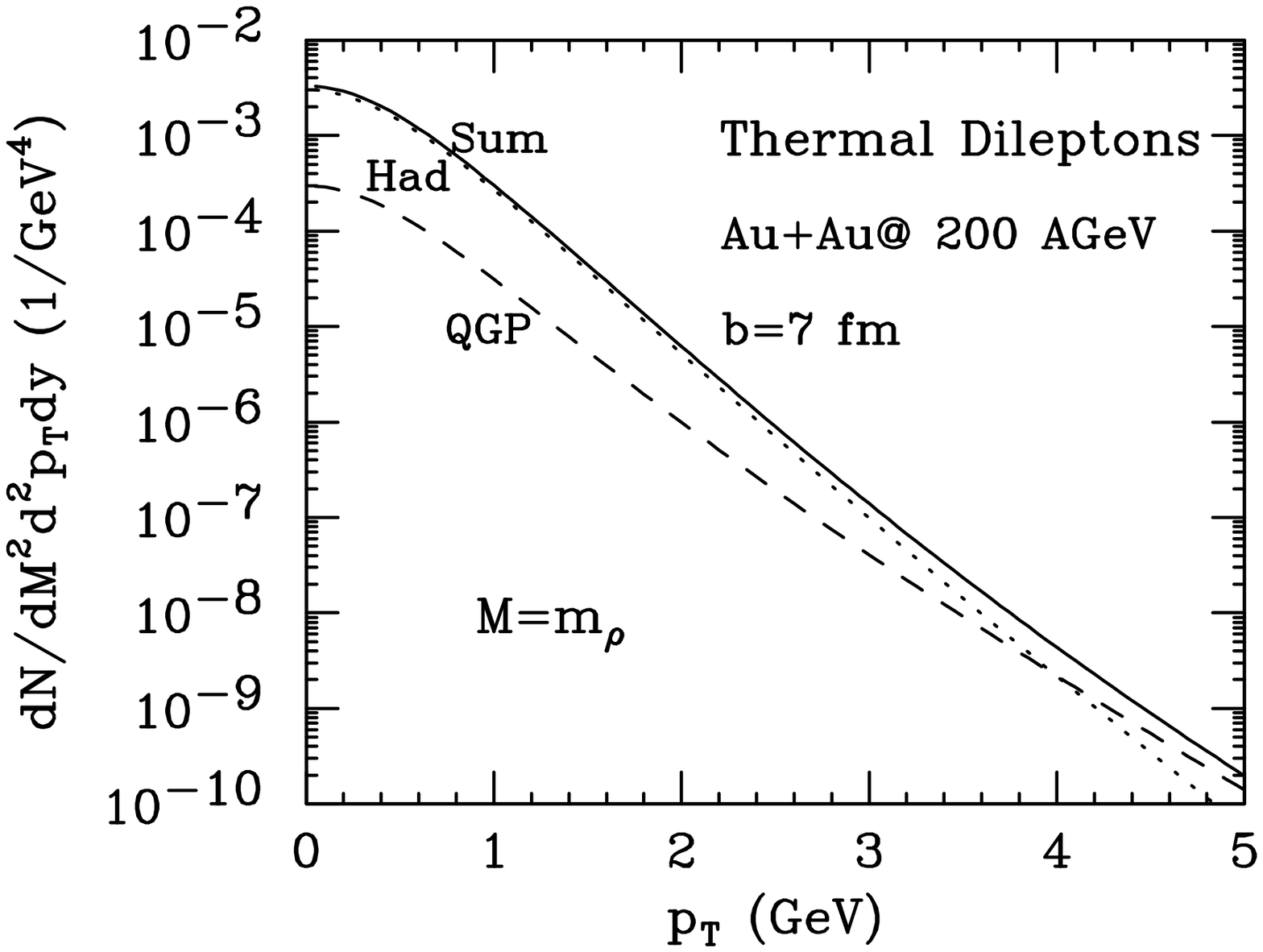,bb=51 208 546 573,width=7.9cm}} 
\centerline{\epsfig{file=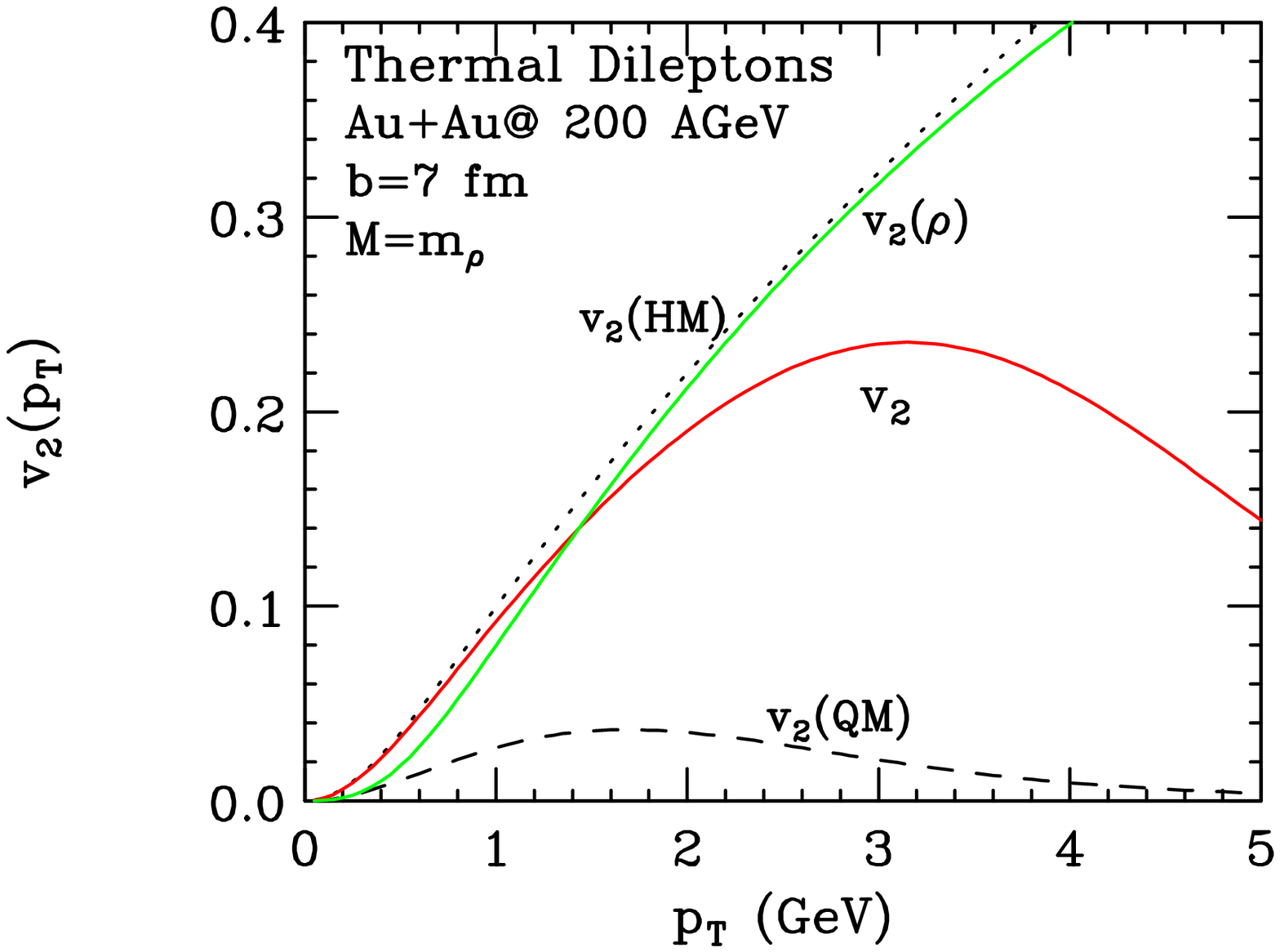,bb=54 208 546 568,width=7.9cm}} 
\caption{(Color online) The transverse momentum spectrum (upper panel)  
and differential elliptic flow $v_2(p_T)$ (lower panel) of thermal  
dileptons with invariant mass $M{\,=\,}m_\rho$. Dashed and dotted lines 
show quark matter (QGP,QM) and hadron matter (Had,HM) contributions,  
respectively. In the bottom panel we also show for comparison the  
elliptic flow of $\rho$ mesons emitted from the hadronic decoupling  
surface.  
\label{fig2}} 
\end{figure} 
%%%%%%%%%%%%%%%%%%%%%%%%%%%%%%%%%%%%%%%%%%%%%%%%%%%%%%%%%%%%%%%%%%%%%%%%%%%% 
% 
 
% 
%%%%%%%%%%%%%%%%%%%%%%%%%%%%%%%%%%%%% Fig. 3 %%%%%%%%%%%%%%%%%%%%%%%%%%%%%%% 
\begin{figure}[t] 
%\centerline{\epsfig{file=dn_phi.ps,bb=51  208  546  564,width=7.9cm}} 
%\centerline{\epsfig{file=v2_phi.ps,bb=54  208  546  568,width=7.9cm}} 
\centerline{\epsfig{file=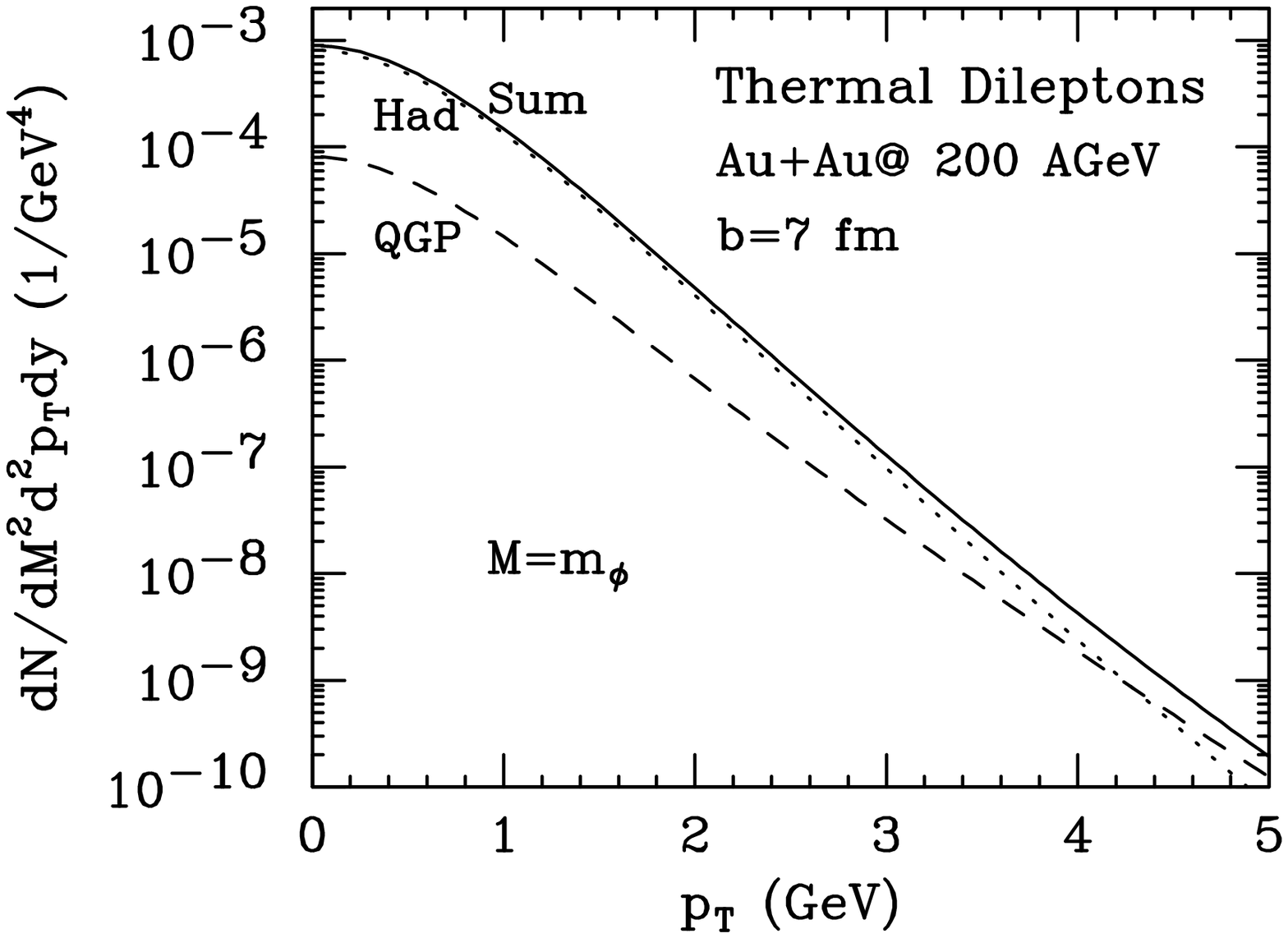,bb=51  208  546  564,width=7.9cm}} 
\centerline{\epsfig{file=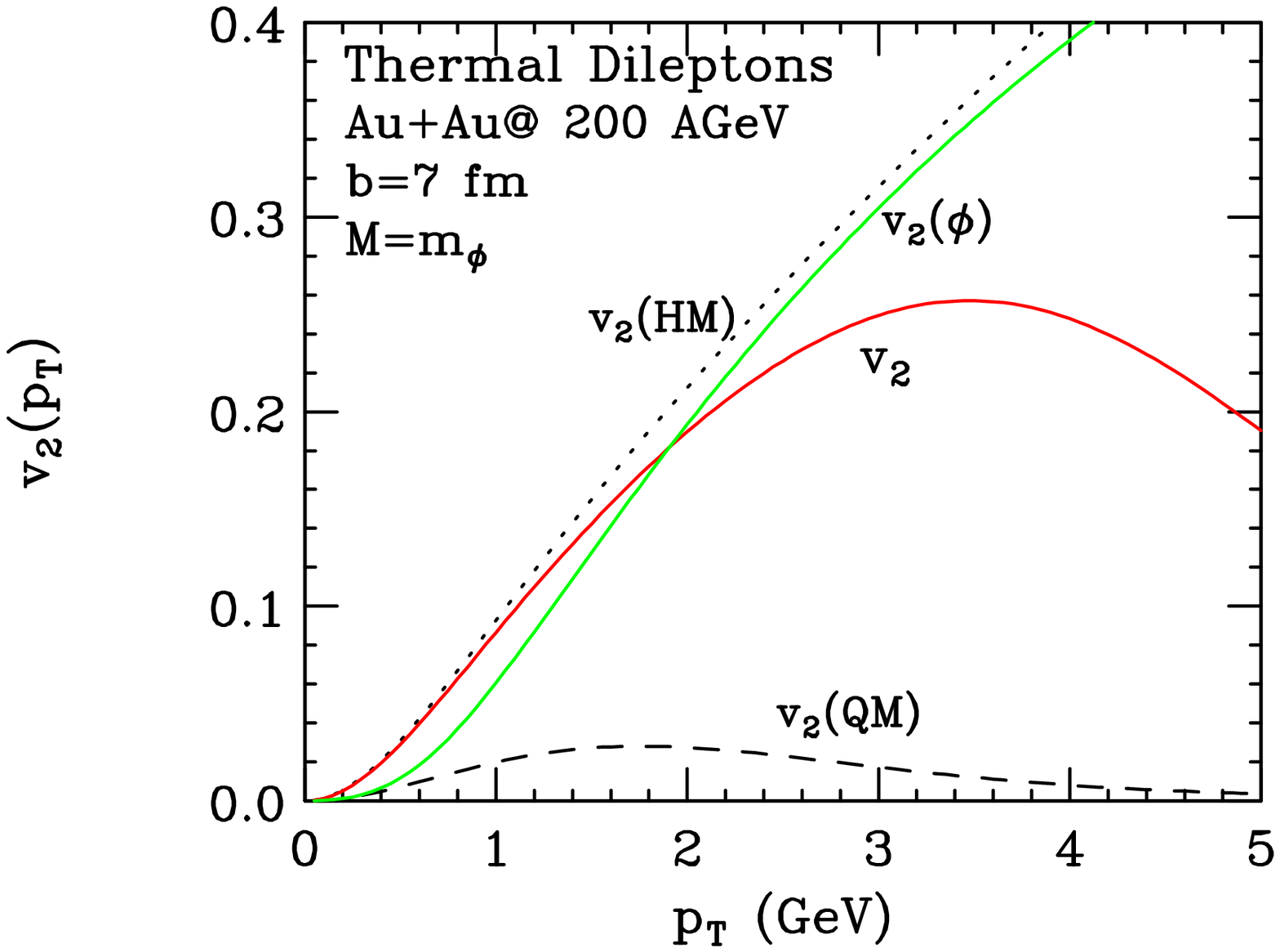,bb=54  208  546  568,width=7.9cm}} 
\caption{(Color online) Same as Fig.~\ref{fig2}, but for dileptons with  
invariant $M=m_\phi$, and with the elliptic flow of $\phi$ mesons from 
the hadronic decoupling surface shown in the bottom panel for comparison. 
\label{fig3}} 
\end{figure} 
%%%%%%%%%%%%%%%%%%%%%%%%%%%%%%%%%%%%%%%%%%%%%%%%%%%%%%%%%%%%%%%%%%%%%%%%%%%% 
% 
We pause here for a moment to note that the relative contributions
of the emission from the hadronic and quark matter stages control
the overall $v_2$ for the dileptons. Thus we shall see in the following 
that the elliptic flow of dileptons with masses near those of the
$\rho$, $\omega$, or $\phi$ mesons is decided by the $v_2$ of the 
radiation from the hadronic matter. The elliptic flow parameter
for dileptons with $M{\,\gg\,}1$\,GeV, on the other hand, is dominated 
by quark matter radiation (see Fig.~\ref{fig6}). To measure this
experimentally would obviously be very valuable.

In Figures~\ref{fig2} and \ref{fig3} we show the transverse momentum  
spectra and $p_T$-dependent elliptic flow $v_2(p_T)$ for dileptons 
with invariant masses $M{\,=\,}m_\rho$ (Fig.~\ref{fig2}) and  
$M{\,=\,}m_\phi$ (Fig.~\ref{fig3}), from semiperipheral Au+Au collisions 
at impact parameter $b{\,=\,}7$\,fm. The bottom panels of these figures 
show that the elliptic flow of hadronic dileptons, $v_2(\mathrm{HM})$, is  
large and basically agrees with the elliptic flow of hadrons of the same  
mass ($v_2(\rho)$ and $v_2(\phi)$, respectively), which are emitted from  
the hadronic freeze-out surface. The somewhat smaller differential elliptic  
flow $v_2(p_T)$ for the hadrons (compared to that of dileptons with the  
same mass) arises from their somewhat flatter $p_T$ spectrum (not shown).  
Since the hadrons decouple later than the average hadronic dilepton, their  
spectra are boosted by somewhat larger radial flow, and flatter spectra  
result in smaller differential elliptic flow $v_2(p_T)$ (see last paper  
in Ref.~\cite{v2_theo}). The elliptic flow of thermal dileptons from the  
QGP phase, $v_2(\mathrm{QM})$, is much smaller and shows the same decrease  
at large $p_T$ that we already observed for thermal photons \cite{CFHS}. It  
also decreases with increasing dilepton mass. These features reflect the  
small flow anisotropies during the early QGP stage, especially for  
dileptons with large $p_T$ and $M$ whose production requires high 
temperatures and is therefore peaked at very early times. One should
keep in mind, however, that the present analysis does not include 
lepton pairs produced via the Drell-Yan and jet fragmentation processes,
nor those produced through jet-plasma interactions \cite{jplas}. These 
can contribute their own characteristic azimuthal asymmetries \cite{TGF}. 
A consistent inclusion of these non-thermal contributions is part of a 
study in progress.  
 
The total dilepton flow $v_2$ is seen to follow that of the hadronic  
dileptons, $v_2(\mathrm{HM})$, up to rather large transverse momenta  
of about 2\,GeV/$c$, before peaking near 3\,GeV/$c$ and then decreasing  
at even larger $p_T$. This is due to the strong dominance of hadronic  
dilepton emission near the vector meson peaks in the dilepton mass  
spectrum of Fig.~\ref{fig1}: As seen in the top panels of Figs.~\ref{fig2}  
and \ref{fig3}, the dilepton $p_T$ spectrum is dominated by hadronic  
emission over almost the entire shown $p_T$ range, with QGP radiation  
winning out only for transverse momenta above $\sim 4$\,GeV/$c$. Note 
that a possible medium-induced suppression of the hadronic dilepton 
contribution by $\sim50\%$ near the $\rho$ mass would decrease the 
net $v_2$ there by only 10\%, because it is a weighted average. Still, 
the absence of non-thermal sources such as those enumerated at the end 
of the preceding paragraph must be kept in mind and should be corrected
before attempting a quantitative comparison with future dilepton flow data.  
 
% 
%%%%%%%%%%%%%%%%%%%%%%%%%%%%%%%%%%%%% Fig. 4 %%%%%%%%%%%%%%%%%%%%%%%%%%%%%%% 
\begin{figure}[ht] 
%\centerline{\epsfig{file=dn_phi_meson.ps,bb=51  208  553  573,width=8cm}} 
\centerline{\epsfig{file=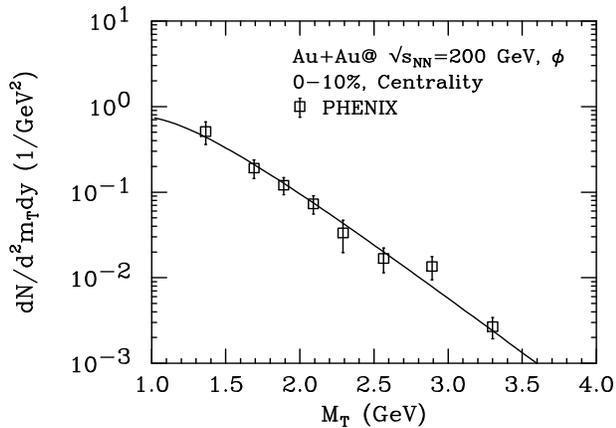,bb=51 208 553 573,width=8cm,height=5.5cm}} 
\caption{Transverse momentum spectrum of $\phi$ mesons,  
comparing the theoretical prediction from the hydrodynamic model 
with PHENIX data Ref.~\cite{phenix_phi}. 
\label{fig4}} 
\end{figure} 
%%%%%%%%%%%%%%%%%%%%%%%%%%%%%%%%%%%%%%%%%%%%%%%%%%%%%%%%%%%%%%%%%%%%%%%%%%%% 
% 
In Figure~\ref{fig4} we show that the success of the hydrodynamic model  
in describing all measured hadron $p_T$ spectra, at least up to  
$p_T{\,\simeq\,}2$\,GeV/$c$, carries over to that of the $\phi$ meson, 
recently measured by the PHENIX Collaboration \cite{phenix_phi} via 
reconstruction of $\phi$ mesons from $K^+K^-$ pairs. It can thus be hoped 
that the predicted spectrum and elliptic flow of dileptons with the same 
mass, as shown in Fig.~\ref{fig3}, will be similarly reliable. 
 
% 
%%%%%%%%%%%%%%%%%%%%%%%%%%%%%%%%%%%%% Fig. 5 %%%%%%%%%%%%%%%%%%%%%%%%%%%%%% 
\begin{figure}[hb] 
%\centerline{\epsfig{file=dn_2.0.ps,bb=51  208  546  573,width=7.9cm}} 
%\centerline{\epsfig{file=v2_2.ps,bb=54  208  546  568,width=7.9cm}} 
\centerline{\epsfig{file=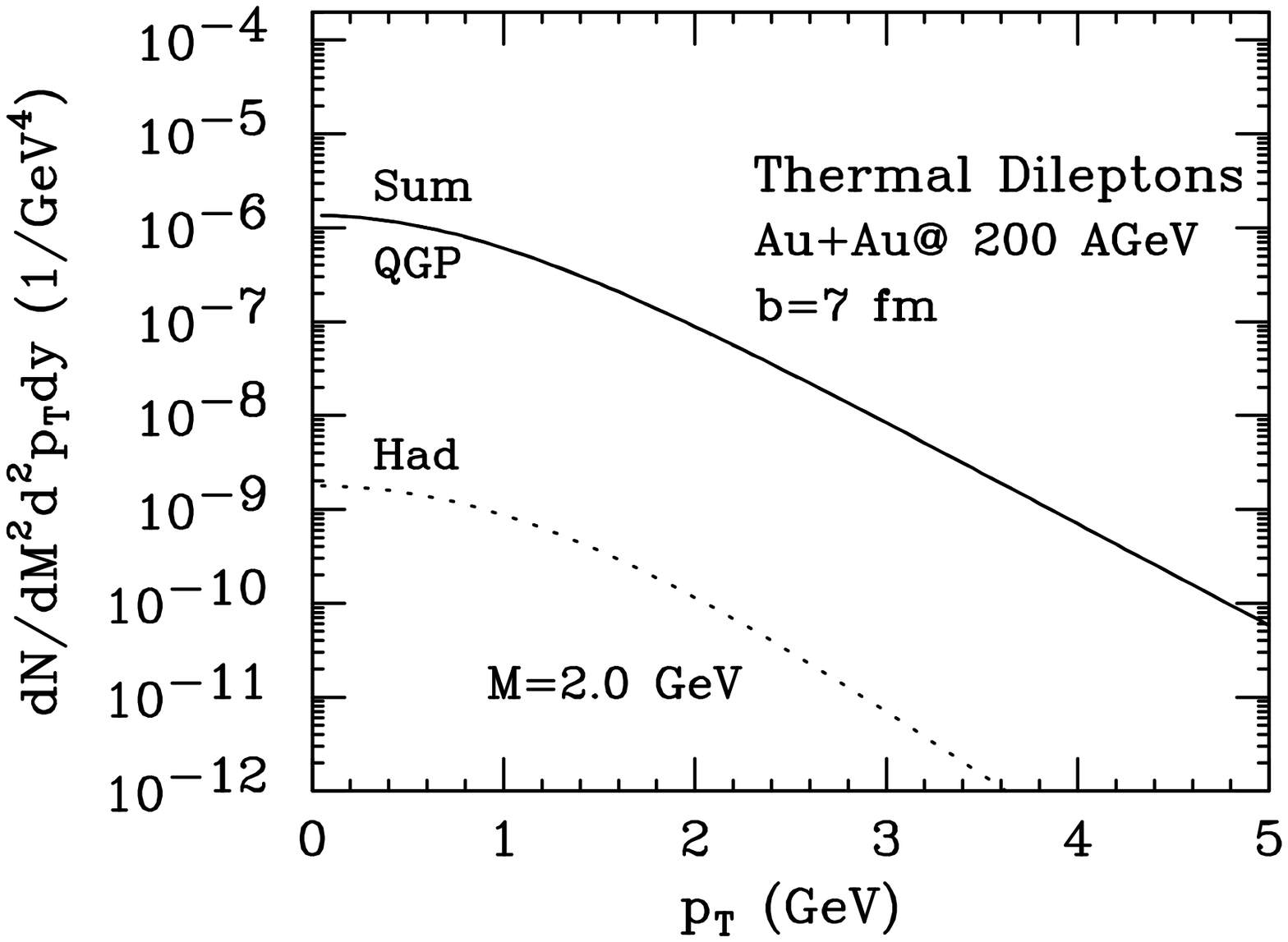,bb=51  208  546  573,width=7.9cm,%
                    height=5.5cm}} 
\centerline{\epsfig{file=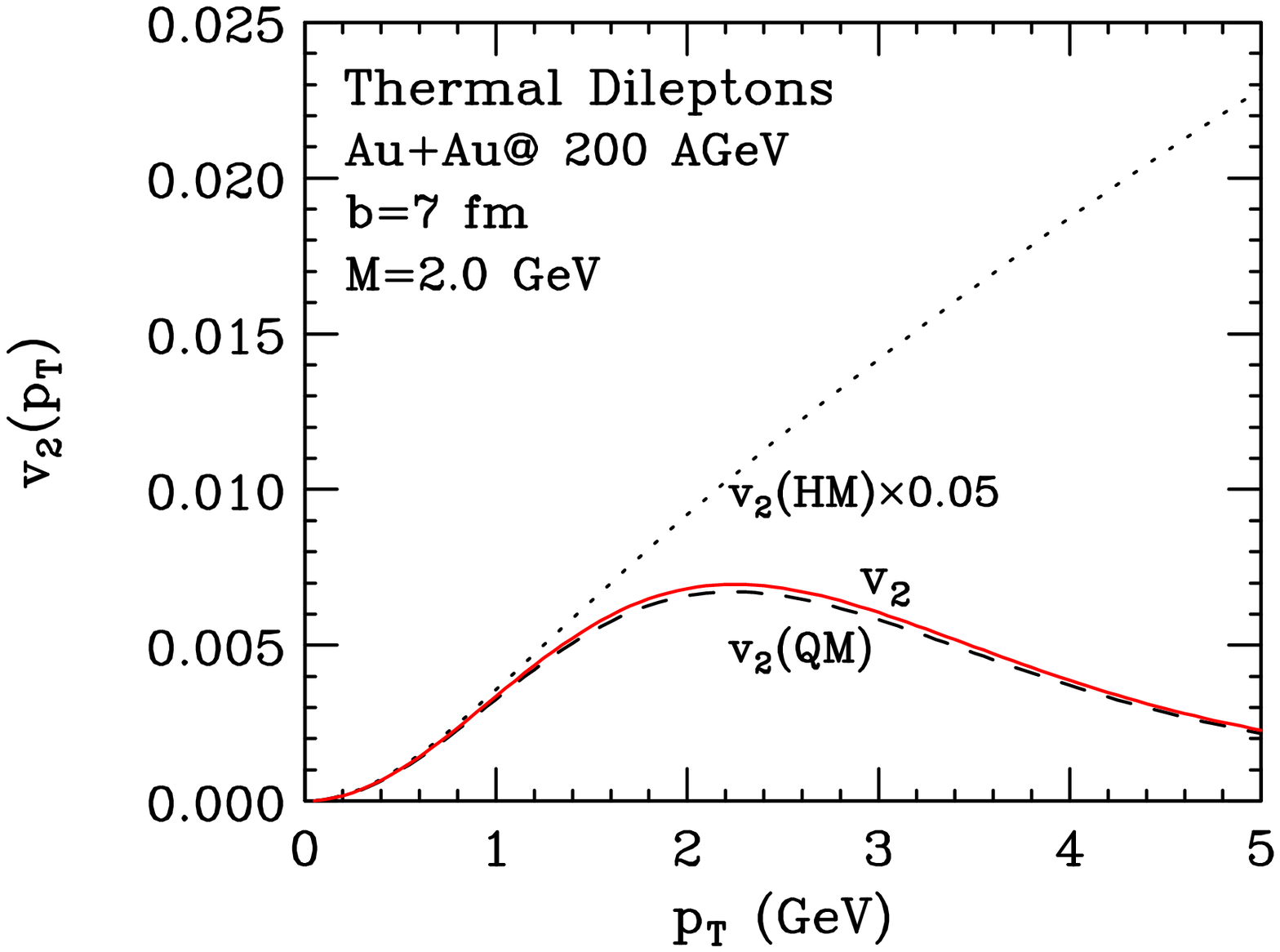,bb=54  208  546  568,width=7.9cm,%
                    height=5.5cm}} 
\caption{(Color online) Same as Fig.~\ref{fig2}, for $M{\,=\,}2$\,GeV.} 
\label{fig5} 
\end{figure} 
%%%%%%%%%%%%%%%%%%%%%%%%%%%%%%%%%%%%%%%%%%%%%%%%%%%%%%%%%%%%%%%%%%%%%%%%%%%% 
% 
For dileptons with invariant mass $M{\,=\,}2$\,GeV the $p_T$ spectrum is 
completely dominated by emission from the QGP (Fig.~\ref{fig5}, top panel). 
Correspondingly, the total dilepton elliptic flow $v_2$ closely follows  
that of the quark matter contribution, $v_2(\mathrm{QM})$, giving us a direct  
measurement of the flow anisotropy developed during the quark phase. Its 
decrease at high $p_T$ again reflects the decreasing flow anisotropies 
at earlier and earlier times. The total elliptic flow is small, peaking  
near $p_T{\,\simeq\,}2$\,GeV/$c$ at a value of about 0.5\%, even though  
the elliptic flow of the (tiny) hadronic contribution of the same mass  
is about 20 times larger.  
 
The upper panel of Figure~\ref{fig6} shows the $p_T$-integrated elliptic 
flow as a function of dilepton mass, $v_2(M)$. The solid line gives the 
total elliptic flow of all dileptons (excluding, however, post-freeze-out 
decay dileptons), while the dashed and dotted lines show the elliptic 
flow of the QGP and hadronic dileptons separately. Note that the latter 
show similar qualitative behaviour as functions of $M$ and of $p_T$: 
while the elliptic flow of hadronic dileptons increases monotonically 
with $M$ and $p_T$, the quark matter dileptons exhibit elliptic flow 
that first rises, then peaks and finally decreases with increasing $M$ 
and/or $p_T$ \cite{fn1}. For comparison, that panel also indicates the 
$p_T$-integrated elliptic flow values for a variety of stable hadron 
species and hadronic resonances emitted from the hadronic decoupling 
surface (see also the last paper in Ref.~\cite{v2_theo}). For large 
invariant masses the $p_T$-integrated elliptic flow of hadronic 
dileptons is seen to approach that of hadrons with the same mass. This 
reflects the fact that the homogeneity regions \cite{fn2} for massive 
particles are small since their thermal wavelength decreases like 
$1/\sqrt{M}$, and therefore all particles of that mass (hadrons or 
virtual photons) feel approximately the same flow. Low-mass hadronic 
dileptons, on the other hand, are emitted from much larger homogeneity 
regions and thus have an appreciable chance of being emitted significantly 
earlier than the corresponding hadrons of the same mass; on average they 
thus carry less radial and elliptic flow. 
 
% 
%%%%%%%%%%%%%%%%%%%%%%%%%%%%%%%%%%%%% Fig. 6 %%%%%%%%%%%%%%%%%%%%%%%%%%%%%% 
\begin{figure}[ht] 
%\centerline{\epsfig{file=v2m_new.ps,width=8.2cm}} 
\centerline{\epsfig{file=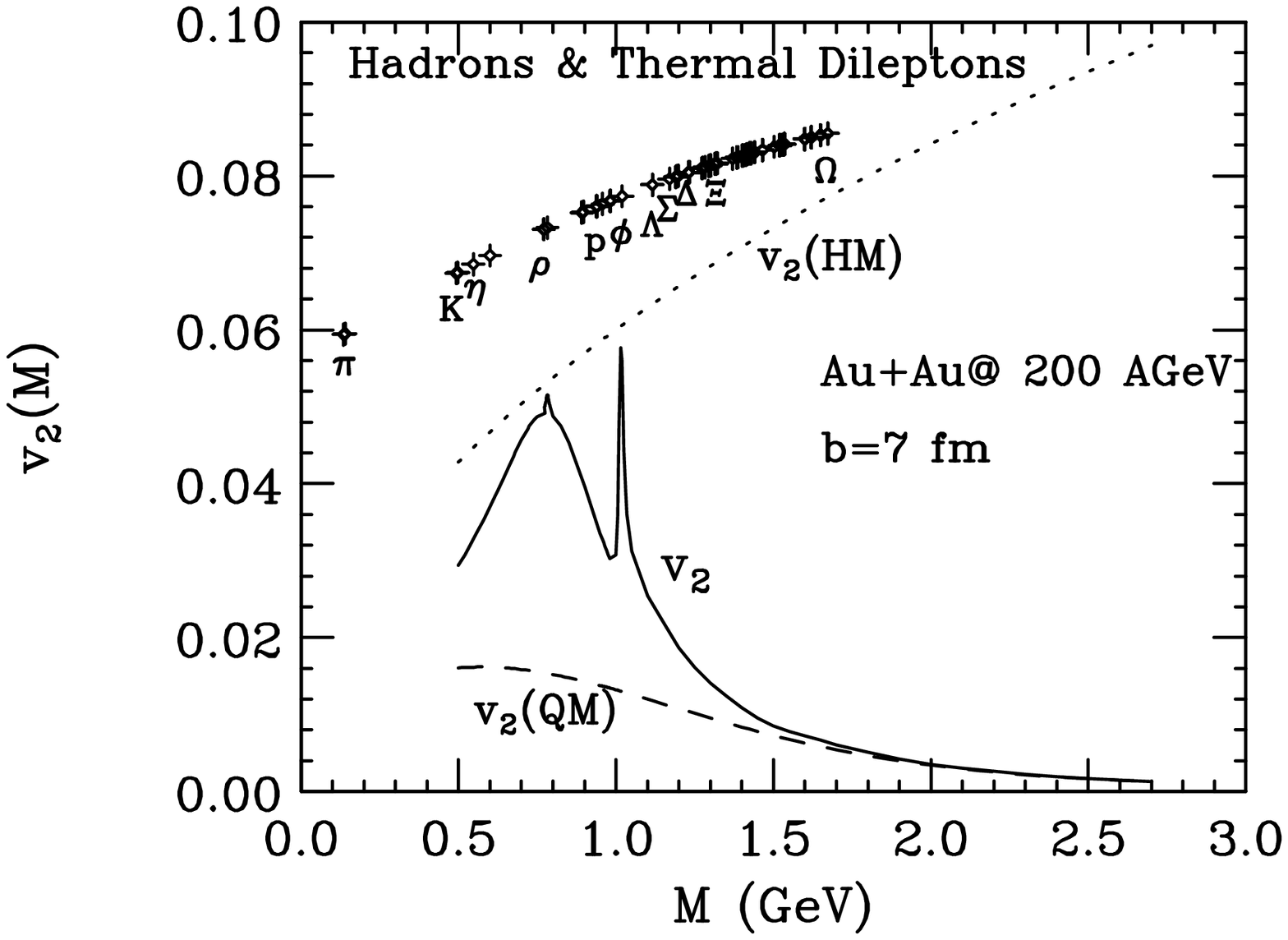,width=8.2cm,height=5.5cm}} 
\centerline{\epsfig{file=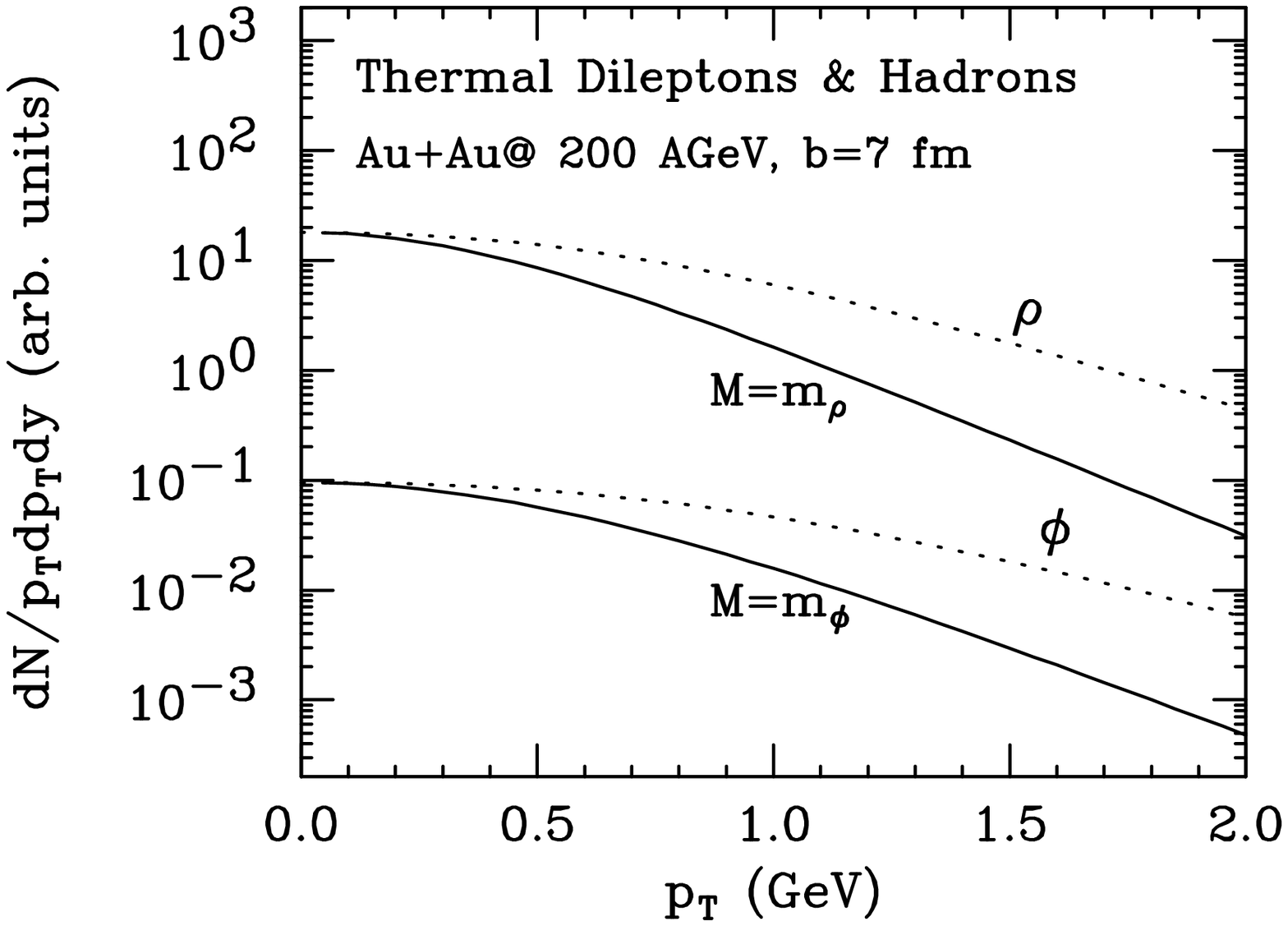,width=8.2cm,height=5.5cm}} 
\centerline{\epsfig{file=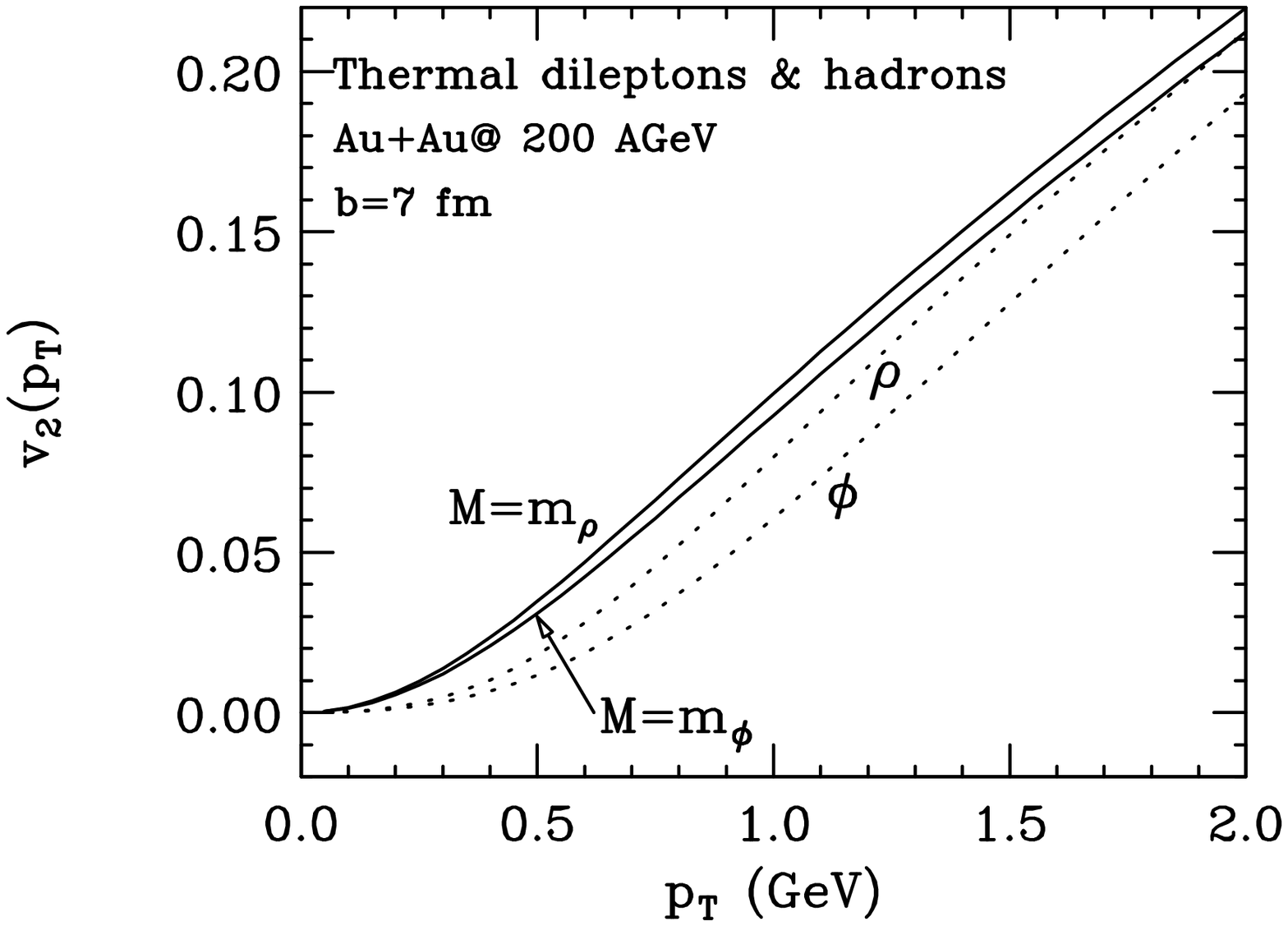,width=8.2cm,height=5.5cm}} 
\caption{
{\sl Upper panel:} $p_T$-integrated elliptic flow parameter for dileptons 
and various hadrons.
{\sl Middle panel:} $p_T$ distributions of dileptons with $M=m_\phi$ 
and $M=m_\rho$, and of $\phi$ and $\rho$ mesons. 
{\sl Lower panel:} Differential elliptic flow for the above.\\[-4mm] 
} 
\label{fig6} 
\end{figure}
\begin{figure}[ht] 
%\centerline{\epsfig{file=v2m_b_new.ps,width=8.2cm}} 
%\centerline{\epsfig{file=v2ecc.eps,width=8.2cm}} 
\centerline{\epsfig{file=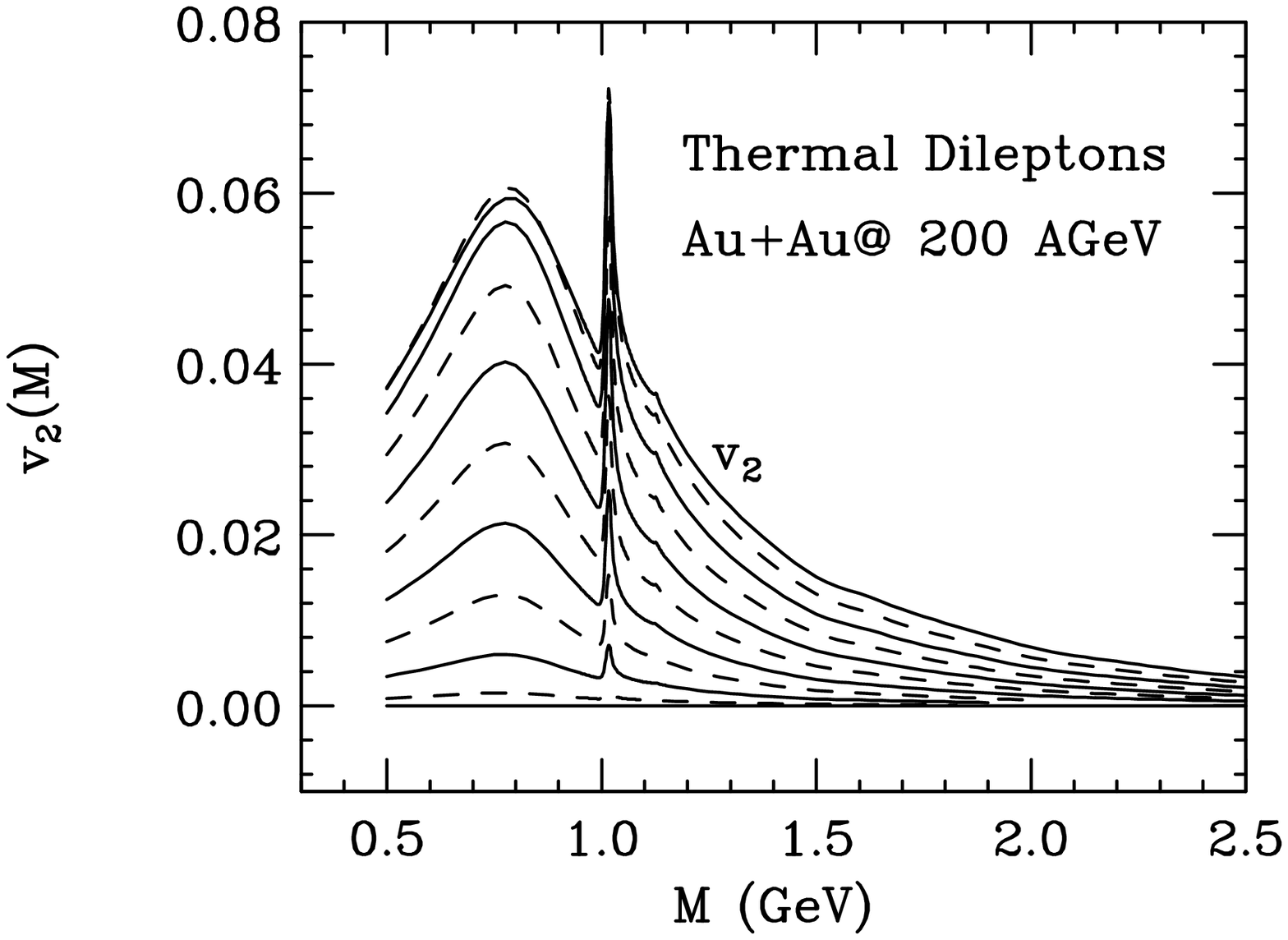,width=8.2cm,height=5.5cm}} 
\centerline{\epsfig{file=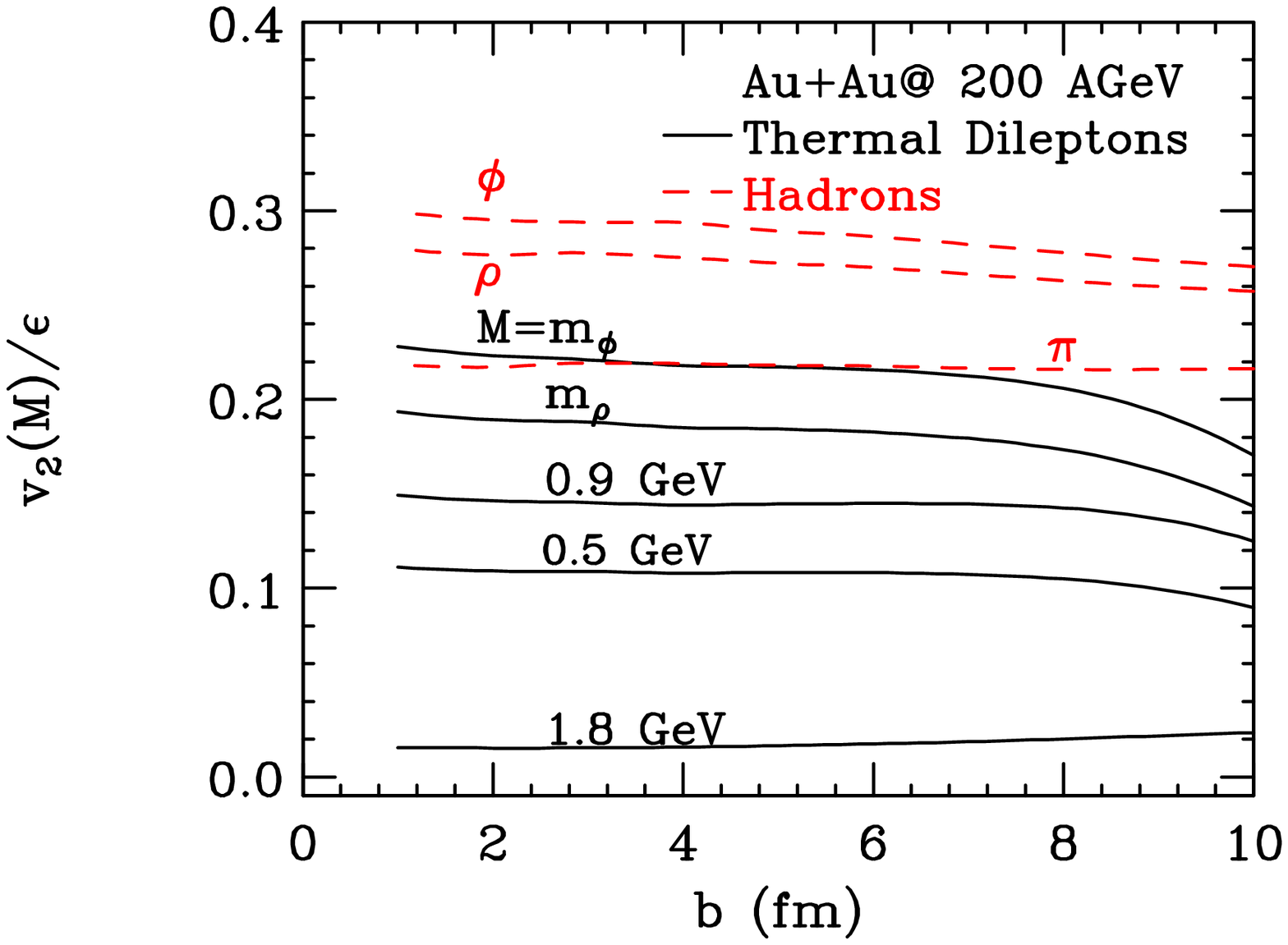,width=8.2cm,height=5.5cm}} 
\caption{(Color online) {\sl Top panel:} 
Dilepton mass dependence of the $p_T$-integrated elliptic flow parameter 
for thermal dileptons, for a variety of impact parameters
($b=$ 0, 2, 4, 6, 8, 10 fm from bottom to top (solid curves) and 
$b=$ 1, 3, 5, 7, 9 fm (dashed curves)). Note that $v_2$ for $b=10$\,fm 
is marginally smaller than for $b=$ 9 fm. 
{\sl Bottom panel:} Impact parameter dependence of the $p_T$-integrated  
elliptic flow parameter scaled by the initial spatial eccentricity 
$\epsilon$, for a variety of hadrons and dileptons with different 
masses.
\label{fig7}} 
\end{figure}

 
%%%%%%%%%%%%%%%%%%%%%%%%%%%%%%%%%%%%%%%%%%%%%%%%%%%%%%%%%%%%%%%%%%%%%%%%%%%% 
% 
%%%%%%%%%%%%%%%%%%%%%%%%%%insertion to replace fn3 %%%%%%%%%%%%%%%%%%%%%%
We note that the additional buildup of radial and (to a lesser extent) 
elliptic flow during the hadronic stage has opposite effects on the 
$p_T$-integrated and $p_T$-differential elliptic flows: While the 
$p_T$-integrated elliptic flow increases, the additional broadening 
of the single-particle spectrum from extra radial flow shifts the 
weight of the flow anisotropy to larger $p_T$, so {\em at fixed $p_T$} 
the elliptic flow decreases (at least in the low-$p_T$ region where 
$v_2(p_T)$ is a rising function of $p_T$). This follows directly from the
discussion presented in the last paper of Ref.~\cite{v2_theo}, and it
is clearly seen in the figures given in the middle and the lower panels 
of Fig.~\ref{fig6}. The middle panel shows the (arbitrarily normalized) 
$p_T$ distributions of dileptons having masses equal to $m_\rho$ and 
$m_\phi$, respectively, compared to those of the $\rho$ and $\phi$ 
mesons themselves. The hadron $p_T$-spectra are clearly seen to decrease
more slowly with $p_T$, as stated above. (We stress that the continued 
hydrodynamical growth of radial flow during the hadronic matter stage 
is essential for this pattern; if one neglects transverse flow one 
arrives at the opposite conclusion \cite{Voloshin}.) The lower panel
in Fig.~\ref{fig6} then demonstrates that this flow-induced flattening 
of the hadronic $p_T$-spectra leads to a systematic lowering of the 
differential elliptic flow $v_2(p_T)$ for these hadrons when compared to
dileptons of similar masses. We shall see later in Fig.~\ref{fig7} 
(lower panel) that this is accompanied by larger values for the 
$p_T$-integrated flow paramters for $\rho$ and $\phi$ mesons
compared to that of dileptons having similar masses.
 
%%%%%%%%%%%%%%%%%%%%%%%%%%%%%%%%%%%%%%%%%%%%%%%%%%%%%%%%%%%%%%%%%%%%%%%%%

As seen in the upper panel of Fig.~\ref{fig6}, the elliptic flow of the 
total dilepton spectrum shows dramatic structure as a function of 
dilepton mass, oscillating between the quark and hadronic matter limits. 
This reflects the relative weight of the QM and HM contributions to the 
dilepton mass spectrum: Close to the vector meson resonance peaks, 
dilepton emission is strongly dominated by emission from the hadron 
matter phase, and the solid line almost reaches the dotted curve for 
$v_2(\mathrm{HM})$. At low dilepton masses, between the $\rho$ and 
$\phi$ mesons, and for dilepton masses $\gtrsim1.5$\,GeV, quark matter 
radiation dominates the dilepton spectrum, resulting in small elliptic  
flow coefficients (of order 1\% or less) for the total dilepton spectrum, 
consistent with the pure quark matter contribution $v_2(\mathrm{QM})$  
(dashed line). By setting appropriate invariant mass windows on the  
dileptons one can thus peek into the early QGP phase and study the  
beginning of elliptic flow buildup.   
 
% 
%%%%%%%%%%%%%%%%%%%%%%%%%%%%%%%%%%%%% Fig. 7 %%%%%%%%%%%%%%%%%%%%%%%%%%%%%% 
%%%%%%%%%%%%%%%%%%%%%%%%%%%%%%%%%%%%%%%%%%%%%%%%%%%%%%%%%%%%%%%%%%%%%%%%%%%% 
%
Figure~\ref{fig7} shows the impact parameter dependence of the  
$p_T$-integrated elliptic flow parameter for thermal dileptons, as a  
function of dilepton mass $M$. The top panel shows that, as the impact 
parameter (and thus the spatial eccentricity 
$\epsilon=\frac{\langle y^2-x^2\rangle}{\langle y^2+x^2\rangle}$ of the 
initial nuclear overlap region) increases, the elliptic flow $v_2$ 
increases, too, as naively expected. In the bottom panel we therefore 
study the eccentricity-scaled elliptic flow, $v_2/\epsilon$, as a function 
of collision centrality, for several hadron species and dileptons with 
a variety of different masses. Even though the dileptons are, on average, 
emitted earlier (i.e. with less radial flow) than the hadrons, we see no 
dramatic differences between the impact parameter dependences of hadronic
and dilepton $v_2$: Both are almost independent of collision centrality,
except for the small drop of $v_2/\epsilon$ for dileptons with masses 
$M\lesssim1$\,GeV at large impact parameters $b\gtrsim 8$\,fm. This 
reaffirms the general understanding \cite{Ollitrault,v2_theo} that for
ideal fluids the ratio $v_2/\epsilon$ reflects the effective stiffness 
(sound speed) of the fireball medium's equation of state, averaged over
the expansion history. The changing weight between QGP and hadronic 
dilepton emission for different dilepton masses is seen to only affect 
the magnitude of their elliptic flow, but not its collision centrality 
(in-)dependence. [Note that the $\omega$ contribution was not included 
in Fig.~\ref{fig7}. The also omitted Drell-Yan contribution is expected 
to emerge only at larger invariant masses than those shown in 
Fig.~\ref{fig7} \cite{RR}.]

%%%%%%%%%%%%%%%%%%%%%%%%%%%%%%%%%%%%%%%%%%%%%%%%%%%%%%%%%%%%%%%%%%%%%%%%%%% 
\section{Conclusions} 
%%%%%%%%%%%%%%%%%%%%%%%%%%%%%%%%%%%%%%%%%%%%%%%%%%%%%%%%%%%%%%%%%%%%%%%%%%% 
 
We have presented a first hydrodynamic calculation of elliptic flow of  
thermal dileptons emitted from ultrarelativistic heavy-ion collisions 
at RHIC energies. The azimuthal flow parameter exhibits a rich structure  
as a function of transverse momentum and invariant mass. When combined  
with photon elliptic flow measurements \cite{CFHS}, this yields a new  
versatile and potentially very powerful probe of the fireball dynamics  
at RHIC and LHC, complementary to the already well-studied flow  
anisotropies in the hadronic sector. $v_2^\gamma(p_T)$ and  
$v_2^{\ell\bar\ell}(M)$ exhibit rich structures which reflect the interplay  
of different emission processes, opening a window on detailed and  
differential information from a variety of different stages of the  
fireball expansion. The elliptic flow of photons emitted from the late  
hadronic stage was seen \cite{CFHS} to track the $v_2$ of the emitting  
hadrons, suggesting the possibility of subtracting the hadronic emission  
contributions from the total photon signal in order to isolate and study  
in greater detail the elliptic flow of early QGP photons. Similarly, we 
saw in the present study that at low ($M{\,<\,}0.5$\,GeV) and high 
($M > 1.5$\,GeV) invariant masses dilepton emission is completely 
dominated by emission from the early quark matter, without need for 
subtracting hadronic contributions. However, at low $M$ experimental 
backgrounds are large while at high $M$ thermal dilepton yields are 
small. Clean elliptic flow measurements of the early QGP stage are 
therefore hard. 
 
Obviously, when compared to hadrons, electromagnetic probes suffer from  
their relatively small production cross sections, so these measurements  
require dedication. However, the first glimpse offered here suggests that  
photon and dilepton elliptic flow have the potential of turning into  
powerful and highly discriminating tools for heavy-ion phenomenology and 
for differential studies of the fireball expansion. Extensive further  
theoretical investigations are therefore warranted, in particular with  
respect to the inclusion of viscous effects at intermediate and high  
transverse momenta.  
 
\begin{acknowledgments}
C.\,G. thanks J\"org Ruppert and Simon Turbide for useful conversations. 
U.\,H. acknowledges the hospitality of the Institute for Nuclear Theory 
at the University of Washington where parts of this work were completed. 
The research reported here was supported in part by the U.S. Department 
of Energy under contract no. DE-FG02-01ER41190, and in part by the Natural 
Sciences and Engineering Research Council of Canada. 
\end{acknowledgments}
 
%%%%%%%%%%%%%%%%%%%%%%%%%% References %%%%%%%%%%%%%%%%%%%%%%%%%%%%%%%%%%%% 
 

\begin{thebibliography}{99}
\bibitem{fn1} 
  For hadrons it is known that the hydrodynamically predicted monotonic  
  rise of $v_2(p_T)$ is broken by viscous effects at large transverse 
  momenta, leading instead to a saturation of $v_2$ for $p_T\ge1.5-2$\,GeV/$c$ 
  for mesons and for $p_T\ge2-2.5$\,GeV/$c$ for baryons \cite{RANP04}. 
  Within the successful quark coalescence model for intermediate-$p_T$ 
  hadron production \cite{coal} this can be interpreted as quark elliptic 
  flow following the hydrodynamic prediction to $p_T\simeq0.75-0.8$\,GeV/$c$, 
  followed by saturation at larger transverse momenta \cite{RANP04}. Since  
  dileptons with large invariant mass $M$ and/or large $p_T$ are formed in  
  collisions involving high-$p_T$ quarks and antiquarks it is likely that  
  the same viscous effects also break the hydrodynamically predicted  
  monotonic rise of the hadronic dilepton elliptic flow seen in all  
  results presented here. Similarly, viscous effects are expected to  
  further reduce the elliptic flow of quark matter dileptons with large  
  invariant masses and/or $p_T$, below the alredy small values shown here.  
  A quantitative assessment of this issue requires hydrodynamic simulations  
  for viscous fluids and will have to await future studies.
\bibitem{fn2} 
  Homogeneity regions are regions of space-time contributing with roughly 
  constant probability to the emission of particles with a given fixed  
  momentum.
\end{thebibliography}
\end{document}